\documentclass[12pt,usenames,dvipsnames]{article}
\usepackage[usenames,dvipsnames]{xcolor}
\usepackage{subcaption}
\usepackage{tikz}
\usepackage{transparent}

\graphicspath{{figures/}}
\usepackage{COBEEpreprint} %

\usepackage{amsmath}
\usepackage{amssymb}
\usepackage{microtype}
\usepackage{mathcomp}
\usepackage{txfonts}
\usepackage{units}

\title
{
    \newline
    LPWAN based IoT Architecture for Distributed Energy Monitoring in Deep Indoor Environments
}

\author[1]{\underline{Christof Röhrig}}
\author[1]{Benz Cramer}

\affil[1]%
{
    \small
    {
        Dortmund University of Applied Sciences and Arts,
        Institute for the Digital Transformation of Application and Living Domains (IDiAL), 
        Otto-Hahn-Str. 23, 44227 Dortmund, Germany \href{mailto:christof.roehrig@fh-dortmund.de}{christof.roehrig@fh-dortmund.de}, 
        \href{https://orcid.org/0000-0002-3286-3703}{\includegraphics[scale=0.5]{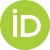}} \hspace{0.8cm} 0000-0002-3286-3703 
    }
}

\usepackage{xfp}

\newcommand{\wisun}{{Wi-SUN\textsuperscript{\tiny \textregistered}}}
\newcommand{\lorawan}{{LoRaWAN\textsuperscript{\tiny \textregistered}}}

\begin{document}
\maketitle              %
\pagestyle{fancy}
\thispagestyle{fancy}
\section*{Summery}
Continuous energy monitoring is essential for identifying potential savings and predicting the energy requirements of buildings.
Energy meters are often located in underground spaces that are difficult to reach with wireless technology.
This paper presents an experimental study comparing different Low Power Wide Area Networks (LPWAN) technologies in terms of building penetration and radio coverage.
The technologies Low Power Long Range Wide Area Networks (\lorawan), Narrow Band Internet of Things (NB-IoT), Sigfox 0G and Wireless Smart Ubiquitous Networks (\wisun) are evaluated experimentally.

It also proposes a distributed hybrid IoT architecture that combines multiple LPWAN technologies using an abstraction layer to optimize cost and coverage.
Communication is message-based using the publish-subscribe messaging pattern. It is implemented using the MQTT protocol.
The abstraction layer decodes the proprietary binary data and converts it to a normalized JSON format.


\section{Introduction}
The Dortmund University of Applied Sciences and Arts operates more than a dozen university buildings spread across the city of Dortmund.
These buildings are heated and air conditioned using various forms of energy: District heating and cooling networks, gas and electricity.
The energy meters are located in underground rooms and energy supply tunnels that are difficult to reach with wireless technology.
Continuous energy monitoring is essential for identifying potential savings and for adaptive model predictive heating control of buildings \parencite{wan:23}.
The paper presents an experimental study that compares different LPWAN (Low Power Wide Area Networks) technologies in terms of building penetration and radio coverage.
These LPWAN technologies differ in terms of topology (star, mesh), power consumption, link budget, radio coverage and radio range.
The experiments will be carried out in various university buildings and in a district heating and cooling tunnel linking several buildings on the campus (see Fig~\ref{fig:EFS}~(a)).

Furthermore, the paper proposes a distributed hybrid IoT architecture that combines different LPWAN technologies to optimize cost and coverage.
The architecture connects existing energy meters and sub-meters with different interfaces (M-Bus, Pulse, SML) via multiple LPWAN radio technologies to the IoT middleware. 

\section{Problem definition and methodical approach}
This research compares four LPWAN radio technologies for smart metering applications in deep indoor and underground environments.
The aim of the research is to investigate the building penetration loss (BPL) of different radio technologies in real-life scenarios.
An empirical evaluation is chosen using a pre-installed radio infrastructure. The measurements are performed with inexpensive
off-the-shelf equipment. We focus on measuring the BPL without considering the power consumption of the equipment.
 
We consider two scenarios: In the first scenario, the radio infrastructure, such as gateways (GWs) and base stations (BSs), is installed inside the building. Radio waves propagate through walls and ceilings. It is assumed that the energy meters are located in the basement of the buildings.
In the second scenario, the wireless infrastructure is installed outdoors. The radio waves propagate through the windows or the exterior wall into the building (Outdoor to Indoor, O2I). Theoretical path loss models for indoor environments and O2I penetration can be found in (\cite{ITU:15}) and (\cite{3GPP:24}) respectively.
It is very difficult to calculate the BPL based only on theoretical models because radio waves propagate in different ways in the building due to reflections and diffraction. Especially in deep indoor and underground scenarios such as basements and tunnels, the real signal path and thus the entire link budget is too complex to be accurately represented by a linear model \parencite{Malarski:19}.

Therefore, we chose an empirical experimental evaluation where the received signal strength indicator (RSSI) is measured at outdoor measurement points (MPs) at ground level and then compared with the RSSI measured at indoor MPs on the first floor and underground MPs in basements and a tunnel (see Fig.~\ref{fig:EFS}~(a)). The BPL is calculated from these measurements and compared between the different radio technologies.

\section{Related Work}
LPWAN technologies have received increased attention in recent years. 
They are characterized by their long range and low power consumption. 
LPWAN technologies known for their good building penetration include \lorawan, NB-IoT, Sigfox 0G and \wisun. 
\textcite{Kadusic:22} provides an overview of \lorawan, NB-IoT and Sigfox and the characteristics of these radio technologies. 
\textcite{Hoo:23} compares \wisun{} with \lorawan{} and NB-IoT. The main application is a building management system.
\textcite{Naumann:21} compares the energy consumption and the link budget of NB-IoT, \lorawan{} and Sigfox based on theoretical models.

\textcite{Roosipuu:23} investigates the use of NB-IoT for monitoring and control of smart urban drainage systems. Different depths of devices in manholes are investigated.  
\textcite{Persia:17} presents a connectivity analysis of NB-IoT and \lorawan{} for smart grid applications in outdoor, indoor and deep indoor environments.
\textcite{Tangsunantham:22} presents an experimental evaluation of \wisun{} for advanced metering in smart grids.
\textcite{Thrane:20} presents an experimental evaluation of an NB-IoT propagation model for deep indoor scenarios. The measurement campaign was conducted in a system of long underground tunnels.

\section{Technology Overview}
LPWAN technologies known for good building penetration operate in the sub-GHz spectrum and include \lorawan, NB-IoT, Sigfox 0G and \wisun. 
\lorawan, Sigfox 0G and \wisun operate in the unlicensed spectrum of \unit[868]{MHz} in Europe, while NB-IoT operates in the licensed spectrum of \unit[800]{MHz} (band 20) or \unit[900]{MHz} (band 8) in Europe. 
To ensure fair use of the unlicensed spectrum, each device is limited in transmit power and duty cycle (DC). 
In the \unit[868]{MHz} band, the transmit power is normally limited to \unit[14]{dBm} (\unit[25]{mW}) and a DC of 0.1\%, 
except in the G1 band (868.0 - \unit[868.6]{MHz}) where a DC of 1\% is allowed and
the G3 band (869.4 - \unit[869.65]{MHz}) where a transmit power of \unit[27]{dBm} (\unit[500]{mW}) and a DC of 10\% is allowed.
The G3 band is typically used for downlink (DL) messages where a BS is connected to many devices and transmits at high power. 
In contrast, quality of service can be guaranteed in the licensed spectrum, where the mobile network operator (MNO) controls the use of the spectrum an the DC is not limited. 

The proprietary Sigfox 0G radio technology was developed by the French company Sigfox S.A. (now owned by UnaBiz). It uses Ultra Narrow-band (UNB), achieves long range and requires low power. 
The Sigfox 0G network is based on a star topology and requires a local 0G network operator to carry the traffic generated.
In Europe, Sigfox 0G is using the G1 band for the UL and the G3 band for the DL.
Sigfox supports up to 140 uplink (UL, from device to BS) and 4 DL messages per day, each carrying a payload of 12 bytes and 8 bytes respectively, at a data rate of \unitfrac[0.1]{kbit}{s}. %
Sigfox 0G subscription cost is 10\,€  per year for one device and 140 UL, 4 DL messages per day (Heliot Europe). 

\lorawan{} is based on the proprietary LoRa radio technology developed by Scemtech.  
\lorawan{} defines the communication protocol and system architecture and is managed by the open LoRa Alliance. 
It can achieve data rates between \unitfrac[0.25]{kbit}{s} and \unitfrac[11]{kbit}{s} in Europe.
The payload of the messages can be from 51 to 222 bytes in Europe depending on the data rate. The number of messages is limited by the DC regulations in Europe and may be even lower due to restrictions by the \lorawan{} network operator. 
The Things Network's (TTN) fair use policy limits a device's airtime to \unit[30]{s} per day. \lorawan{} networks can be operated by a network operator, a community such as TTN, or self-deployed.  
\lorawan{} thus offers a high degree of flexibility for the end user. 

\wisun{} is based on the open IEEE 802.15.4g/e standards. Unlike other LPWAN technologies, it offers mesh and multi-hop features 
for enhanced reliability and range. In addition to low power consumption and long range, it offers lower latency and higher data rate up to \unitfrac[300]{kbit}{s} compared to the other LPWAN technologies. The downside of \wisun{} is the lower link budget and therefore lower range and building penetration, and the lack of off-the-shelf devices for the end user.
The target applications of \wisun{} are in the field of building automation \parencite{Hoo:23}.

NB-IoT operates in licensed spectrum and is standardized by the 3rd Generation Partnership Project (3GPP) as LTE Cat-NB1 and -NB2.
It is only available through MNOs. Implementing NB-IoT is cost effective as it is based on existing cellular infrastructure.
Compared to other LPWAN technologies that operate in unlicensed spectrum, the transmit power and DC are not limited by regulation.
Therefore, the link budget is higher, resulting in better building penetration. The downside is higher power consumption.
NB-IoT offers data rates of up to \unitfrac[62]{kbit}{s} in Cat NB1 and up to \unitfrac[159]{kbit}{s} in Cat NB2 in the UL 
and up to \unitfrac[26]{kbit}{s} and \unitfrac[127]{kbit}{s} respectively in the DL.  
Applications are typically low throughput, delay tolerant and low mobility. Examples are smart meters and remote sensors.
The cost of the subscription is 11\,€ for a SIM card, \unit[500]{MByte} for 10 years (1NCE).

Table~\ref{tab:lpwan} compares the key parameters of the technologies. The link budget is the maximum possible path loss from transmitter to receiver. A high value is important for penetrating underground environments. The values are taken from \parencite{Naumann:21} and the data sheets of the experimental equipment in Table~\ref{tab:dev}.
\begin{table}[h!]
\caption{Comparison of different LPWAN technologies (Europe)}
\label{tab:lpwan}
\begin{center}
{\small
\begin{tabular}{| l | c | c | c | c |}
\hline
Technology                   & \lorawan  & Sigfox  & NB-IoT & \wisun  \\\hline
Topology                     & star     & star    & star   & mesh    \\\hline
Band in MHz                  & 868      & 868     &  800 / 900  &  868 \\\hline
Tx power GW in dBm           & 14 / 27$^*$ & 27    & 23     & 14       \\\hline
Tx power device  in dBm      & 14       & 14      & 23     & 14       \\\hline
Rx sensitivity GW in dBm     & -140     & -144    & -131   & -120     \\\hline 
Rx sensitivity device in dBm & -136     & -132    & -131   & -120     \\\hline
link budget$^+$ UL in dB        & 154      & 158     & 154    & 134      \\\hline
link budget$^+$ DL in dB       & 150 / 163$^*$& 159    & 154    & 134      \\\hline
price of module              & 9€       & 9€      & 12€    & 20€     \\\hline
annual subscription cost     & -        & 10€     & 1-2€   & - \\\hline
Device availability          & ++       & 0       & +      & - -        \\\hline
\end{tabular}
\newline {\tiny $^*$ depending on the frequency of use, $^+$ without antenna gain (0dBi)}
}
\end{center}
\end{table}
\section{Experimental evaluation of LPWAN technologies}
\subsection{Experimental setup}
An area of the campus consisting of several challenging buildings for building penetration was chosen as the experimental environment (see Fig~\ref{fig:EFS}~(a)).
We use several off-the-shelf devices for the experiments (see Table~\ref{tab:dev}). 
Since there are no end-user \wisun{} devices available, we use a development kit from Silicon Labs.
This equipment does not provide an accurate measurement of RSSI, but is intended to provide an estimate of building penetration with real devices.  
\begin{table}[h!]
\caption{Devices used in experiments}
\label{tab:dev}
\begin{center}
{\small
\begin{tabular}{| l | c | c | c | c | c | c |}
\hline
Device              & LPS8 & DLOS8 & LoRa E5 mini& SIM7020E & MKR FOX & SLWSTK6007A \\\hline %
\hline
type                & GW      & GW      & device   & device    & device      & dev kit \\\hline
manufacturer        & Dragino & Dragino & Seeed    & Simcom    & Arduino     & Silicon Labs \\\hline
radio chip          & SX1308  & SX1301  & STM32WLE & MT2625 & ATAB8520E   & EFR32MG12  \\\hline %
technology          & LoRa  & LoRa & LoRa  & NB-IoT    & Sigfox      & \wisun  \\\hline
band MHz         & 868     & 868     & 868      &  900 / 800  & 868         &  868 \\\hline
Tx pow. / dBm      & 14 / 20  & 14 / 27   & 14    & 23      & 13     & 14       \\\hline
Rx sen. / dBm     & -140    & -140    & -136.5   & -131    & -132   & -120     \\\hline 
\end{tabular}
}
\end{center}
\end{table}

\subsection{Indoor experimental evaluation of \lorawan{} and \wisun}
In the first measurement campaign, we compare \lorawan{} with \wisun. The experiments are performed in building 2, where the \lorawan{} gateway GW2 (LPS8) is placed on the first floor near a window.
Measurements are made on different floors and in the basement to measure the penetration loss of the ceilings.
With \lorawan{} it is possible to reach every floor of the building up to the 3rd floor and even the basement.
Penetration loss is ~\unit[10]{dB} for each ceiling upstairs and ~\unit[55]{dB} to the basement.

With \wisun{} it was only possible to penetrate two ceilings. The basement could not be reached with this technology.
In general, it was not possible to establish a stable connection when the RSSI was lower than \unit[-95]{dBm}.
Due to the relatively low link budget, building penetration is not \wisun's strength. Rather, its strengths lie in high data throughput and low latency.

\subsection{O2I penetration of \lorawan, Sigfox OG and NB-IoT}
\begin{figure}[!h]
\begin{center}
\begin{tabular}{c c}
\includegraphics[height=6cm]{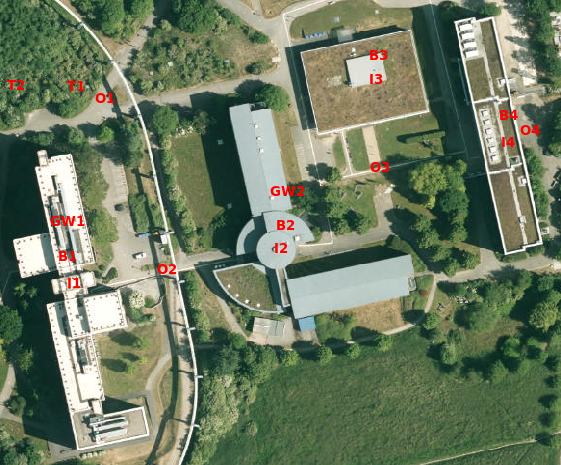} &
\includegraphics[height=6cm]{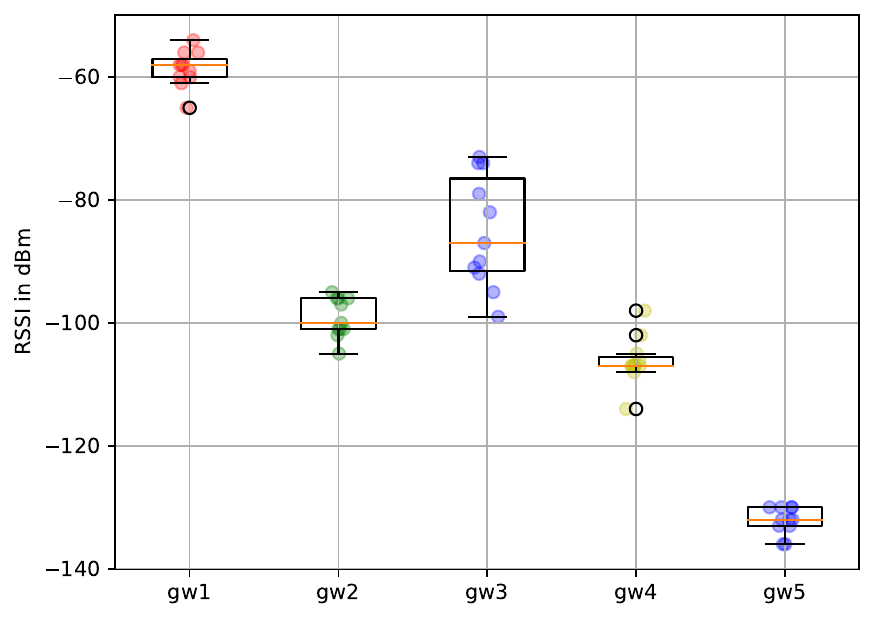}\\
{\footnotesize(a)} & {\footnotesize(b)} \\
\end{tabular}
\end{center}
\caption{(a) Experimental area at campus Emil-Figge-Str. with measurement points marked in red\\ (b) \lorawan RSSI values of Gateways at measurement point O2}
\label{fig:EFS}
\end{figure}

Fig.~\ref{fig:EFS}~(a) shows the experimental area at the campus Emil-Figge-Str. in Dortmund. The outdoor MPs are named O1, O2, O3 and are located between the buildings.
The indoor MPs I1-I4 are located on the first floors in the entrance halls of the buildings. The MPs in the basements of the buildings are labeled B1-B4.
There are two MPs in the energy supply tunnel, T1 and T2. T1 is placed under an emergency manhole of the tunnel, T2 is placed directly at the heat meters.
Two \lorawan{} gateways are placed in the area. The outdoor gateway GW1 (DLOS8) is placed on the roof of building 1, while the indoor gateway GW2 (LPS8) is placed inside on the first floor near a window. Both gateways are preinstalled and not placed for this work.

At each MP, several measurements are made for each radio technology except \wisun.
The results are shown in Table~\ref{tab:RSSI}. 
For \lorawan, the reported RSSI fluctuates a lot. Fig.~\ref{fig:EFS}~(b) shows the fluctuation of the RSSI reported by the gateways.
The values are not normally distributed and have outliers. The RSSI values in the figure are reported by four gateways receiving the messages from a device at O2.  
The RSSI values in the Table~\ref{tab:RSSI} are the medians of the measurements at the indicated MP. 
In this challenging environment, Sigfox 0G was unable to reach the basements of the buildings (except B2 and B4) and the energy supply tunnel. 
\lorawan{} can reach the basements and the tunnel except for MP T2, which has a greater distance to the emergency manhole than T1.
Two MNOs with different frequency bands are evaluated. Vodafone (NB Vo) uses band 20 (\unit[800]{MHz}) for NB-IoT, while Deutsche Telekom (NB DT) uses band 8 (\unit[900]{MHz}).
The RSSI values for NB-IoT are measured DL and are obtained directly from the NB-IoT modem using AT commands. 
The UL RSSI values for \lorawan{} are obtained from the TTN back-end. The DL RSSI is obtained from the device using AT commands.
The Sigfox 0G RSSI values are measures UL at the BS and reported by the Sigfox back-end system. 

\begin{table}[h!]
\caption{RSSI values in dBm for \lorawan{} (LW), NB-IoT (NB) and Sigfox}
\label{tab:RSSI}
\begin{center}
{\small
\begin{tabular}{| c | c | c | c | c | c |}
\hline
MP    & LW UL      & LW DL         & NB DT & NB Vo  & Sigfox   \\\hline
O1    &    -72     &    -68.5      &  -66  & -77     & -96.5 \\\hline %
O2    &    -58     &    -61        &  -66  & -74     & -94\\\hline %
O3    &    -57     &    -55        &  -63  & -73     & -101\\\hline %
O4    &    -83.5   &    -84        &  -64  & -74     & -100\\\hline 
I1    &    -81     &    -78.5      &  -74  & -84.5   & -111\\\hline %
I2    &    -87     &    -90        &  -85  & -90     & -116.5\\\hline %
I3    &    -102    &    -102.5     &  -90  & -99     & -127\\\hline %
I4    &    -93     &    -94        &  -77  & -81     & -114\\\hline %
B1    &    -101    &    -106       &  -95 & -116    & - \\\hline %
B2    &    -111    &    -110       &  -97 & -112    & -137 \\\hline %
B3    &    -114    &    -115       &  -110 & -111    & - \\\hline %
B4    &    -111    &    -106       &  -85  & -94     & -124 \\\hline %
T1    &    -116    &    -112       &  -107 & -114    & - \\\hline %
T2    &       -    &      -        &  -112 & -118    & - \\\hline %
\end{tabular}
}
\end{center}
\end{table}

To compare the BPL of different LPWAN technologies, the RSSI of the indoor MPs is subtracted from the RSSI of the nearest outdoor MP.
Table~\ref{tab:bpl} compares the BPL for each indoor MP. 
\newcommand{\fp}[1]{\fpeval{#1}}
\begin{table}[h!]
\caption{Calculated O2I BPL from nearest outdoor MP to indoor MP in dB}
\label{tab:bpl}
\begin{center}
{\small
\begin{tabular}{| l | c | c | c | c | c | c | c | c |c | c |}
\hline
             & ~I1~         & ~B1~        & ~I2~       & ~B2~        & ~I3~          & ~B3~        & ~I4~       & ~B4~        & ~T1~        & ~T2~ \\\hline
\lorawan{} UL& \fp{81-58}   & \fp{101-58} & \fp{87-58} & \fp{111-57} & \fp{102-57}   & \fp{114-57} & \fp{93-83.5} & \fp{111-83.5} & \fp{116-72} & -\\\hline
\lorawan{} DL& \fp{78.5-61} & \fp{106-61} & \fp{90-61} & \fp{110-55} & \fp{102.5-55} & \fp{115-55} & \fp{94-84} & \fp{106-84} & \fp{112-68.5} & -\\\hline
Sigfox 0G    & \fp{111-96.5}& -           & \fp{116.5 -94} & \fp{137-94} & \fp{127-101}  & -       & \fp{114-101}&  \fp{124-101} & - & - \\\hline
NB-IoT DT B8 & \fp{74-66}   & \fp{95-66} & \fp{85-63} & \fp{97-63} & \fp{90-63}    & \fp{110-63} & \fp{77-64} & \fp{85-64}  & \fp{107-66} & \fp{114-66}\\\hline
NB-IoT Vo B20& \fp{84.5-74} & \fp{116-74} & \fp{90-73} & \fp{112-73} & \fp{99-73}   & \fp{111-73} & \fp{81-74} & \fp{94-74} & \fp{116-77} & \fp{122-77}\\\hline
\end{tabular}
}
\end{center}
\end{table}

Due to measurement inaccuracies, these values are not exact and it is not possible to rank the performance of the technologies in terms of BPL.
In addition, the \lorawan{} GWs adjust the transmit power depending on the RSSI, so the BPL in the DL may be too low. 

The BPL depends more on the signal path than on the LPWAN technology used.
The \lorawan{} GW1 is placed on top of building 1, the signal has to travel through all ceilings to reach the first floor and the basement.
Therefore the BPL of I1 and B1 is much higher for \lorawan compared to NB-IoT and Sigfox 0G.
The basement of building 4 is underground on the west side and above ground on the east side. The signal of \lorawan{} comes from the west, 
that of NB-IoT and Sigfox 0G come from the east, so the BPL of \lorawan{} is much larger than that of NB-IoT and Sigfox 0G.
The Sigfox 0G BS is much further away than the \lorawan{} GWs. 
The angle of entry into the buildings is therefore significantly flatter compared to \lorawan{} and the BPL is in a similar range to NB-IoT.

From the measured values it can be concluded that the angle at which the signal enters the building, and therefore the position of the GW and BS, 
has a greater influence on BPL than the technology used. %
This means that if the LPWAN infrastructure is unknown, the suitability of a LPWAN technology must be determined by measurements.
However, the measured values can be used to determine the order of magnitude of BPL to be expected.
It is clear to see that the MPs in the basements have a very high BPL of up to \unit[60]{dB} for \lorawan, which means that the signal is attenuated by a factor of $10^6$. %

\section{LPWAN architecture}
There are many reasons to choose a LPWAN technology. In addition to range and building penetration, cost, availability of devices and infrastructure play a major role. 
In addition, technologies are constantly evolving. Technologies such as \wisun, which are not widely used today,
may become more prevalent in the future, while other technologies such as Sigfox may become less important.
In particular, the market for LPWAN energy monitoring devices is changing rapidly.
Even if a separate \lorawan{} infrastructure is operated, it is usually cheaper to use an NB-IoT device than to install a separate \lorawan{} gateway for a single device 
if there is no \lorawan{} coverage at the device's location.
\begin{figure}[h!]
\begin{center}
\includegraphics[width=0.80\textwidth]{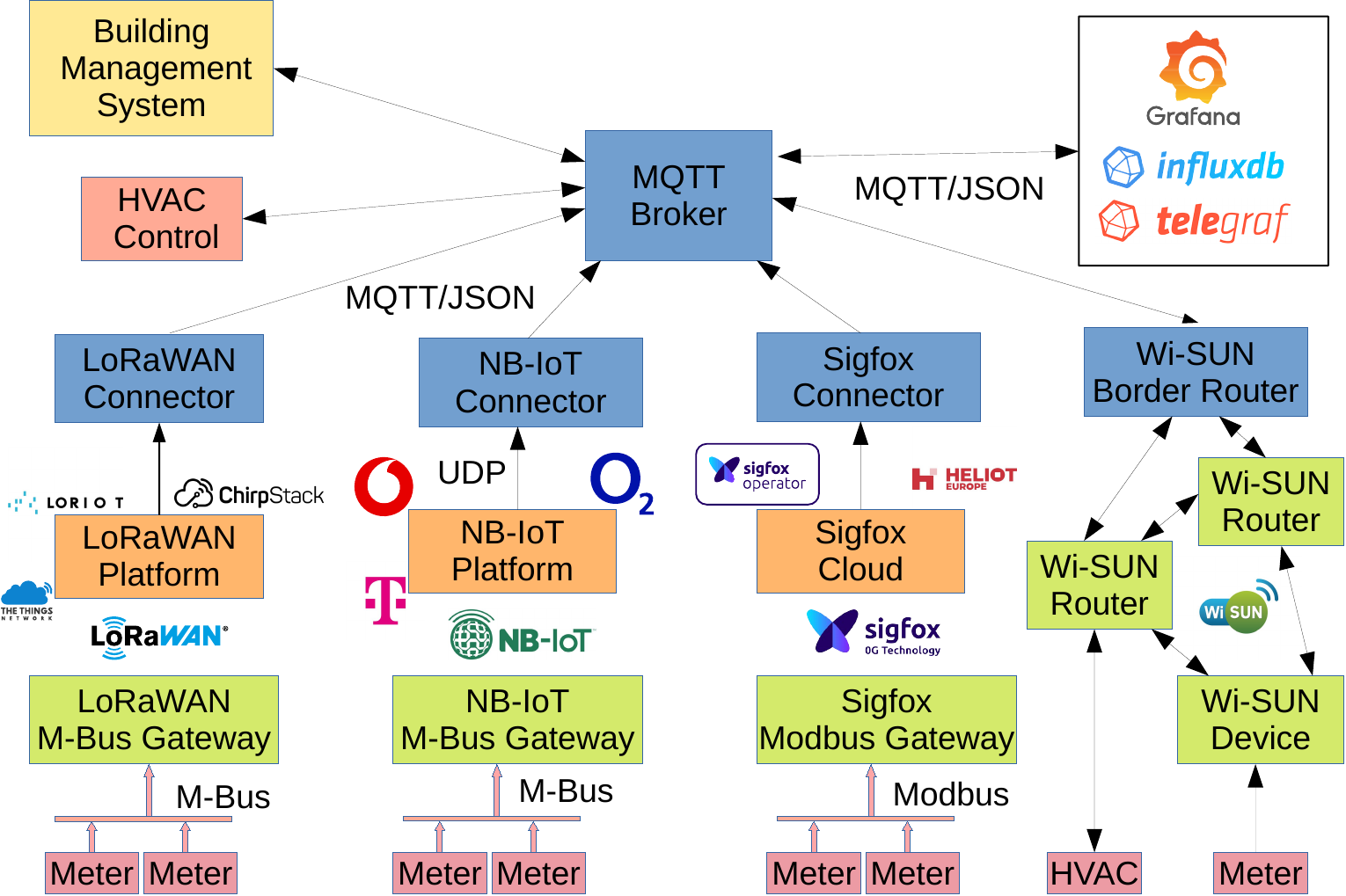}
\end{center}
\caption{\label{fig:arch} Proposed architecture for energy monitoring and control} 
\end{figure}
For this reason, a hybrid and modular architecture is proposed in which multiple LPWAN technologies can be integrated (see Fig.~\ref{fig:arch}). 
The communication is message based and uses the publish-subscribe messaging pattern. It is implemented using the MQTT protocol.
Multiple devices such as meters are connected to a MQTT broker via software connectors.
The software connectors provide an abstraction layer for LPWAN technologies with a common interface based on MQTT and JSON. 

The architecture connects existing energy meters and sub-meters with different interfaces to the IoT middleware via multiple LPWAN radio technologies. 
The district heating and cooling meters provide an M-Bus interface, while the gas meters provide only pulses to be counted.
Energy data from different sources and technologies is transmitted over different wireless technologies. 
A connector for each technology receives the energy data over the Internet, decodes the binary data, and converts it to a normalized JSON format. The energy data is then published to a MQTT broker, which acts as a central hub in the system.
The data is stored in a time series database for later analysis and visualization in a web front-end.

\section{Conclusions}
The paper presents an experimental evaluation of four LPWAN technologies in deep indoor environments. 
\lorawan{} and NB-IoT show better building penetration than Sigfox 0G and \wisun.
By using licensed radio bands and automatic retransmission in case of packet loss, NB-IoT offers higher reliability compared to other LPWAN technologies.
Compared to NB-IoT, \lorawan{} has the advantage of lower power consumption, greater flexibility in the choice of infrastructure, and a wider range of devices available for energy monitoring.
\wisun{} may become more important in the future due to its mesh topology. However, there are currently no energy monitoring devices available for use in Europe.
Furthermore, \wisun{} has the lowest building penetration compared to the other LPWAN technologies considered.
Sigfox 0G generally offers a high link budget and therefore the potential for high building penetration, but only two out of four basements are reached in the experimental evaluation.
This is likely due to the low density of Sigfox BSs in the area studied.

A hybrid and modular architecture for energy monitoring has been proposed that can integrate multiple LPWAN technologies.
The architecture has been implemented and tested for \lorawan{} and NB-IoT.
It is successfully used for energy monitoring in several buildings at the Dortmund University of Applied Sciences and Arts.

\printbibliography[title={\MakeUppercase{\footnotesize References}}]

@Article{Roosipuu:23,
  author   = {Roosipuu, Priit and Annus, Ivar and Kuusik, Alar and Kändler, Nils and Alam, Muhammad Mahtab},
  title    = {Monitoring and control of smart urban drainage systems using NB-IoT cellular sensor networks},
  journal  = {Water Science and Technology},
  year     = {2023},
  volume   = {88},
  number   = {2},
  pages    = {339-354},
  month    = {07},
  issn     = {0273-1223},
  doi      = {10.2166/wst.2023.222},
  url      = {https://doi.org/10.2166/wst.2023.222},
}

@InProceedings{Persia:17,
  author    = {Persia, Samuela and Carciofi, Claudia and Faccioli, Manuel},
  title     = {{NB-IoT} and {LoRA} connectivity analysis for {M2M/IoT} smart grids applications},
  booktitle = {2017 AEIT International Annual Conference},
  year      = {2017},
  pages     = {1-6},
  doi       = {10.23919/AEIT.2017.8240558},
}

@Misc{Naumann:21,
  author = {Naumann, Harald and Oelers, Wilhelm},
  title  = {{LPWAN} Comparison -- Low Energy Consumption with {NB-IoT}, {LoRaWAN} and {Sigfox}},
  year   = {2021},
}

@InProceedings{Thrane:20,
  author    = {Thrane, Jakob and Malarski, Krzysztof Mateusz and Christiansen, Henrik Lehrmann and Ruepp, Sarah},
  title     = {Experimental Evaluation of Empirical NB-IoT Propagation Modelling in a Deep-Indoor Scenario},
  booktitle = {GLOBECOM 2020 - 2020 IEEE Global Communications Conference},
  year      = {2020},
  pages     = {1-6},
  doi       = {10.1109/GLOBECOM42002.2020.9322360},
}

@InProceedings{Kadusic:22,
  author    = {Kadusic, Esad and Ruland, Christoph and Hadzajlic, Narcisa and Zivic, Natasa},
  title     = {The factors for choosing among {NB-IoT}, {LoRaWAN}, and {Sigfox} radio communication technologies for {IoT} networking},
  booktitle = {2022 International Conference on Connected Systems \& Intelligence (CSI)},
  year      = {2022},
  pages     = {1-5},
  doi       = {10.1109/CSI54720.2022.9923999},
}

@Article{3GPP:24,
  author  = {3GPP},
  title   = {Study on channel model for frequencies from 0.5 to 100 {GHz}},
  journal = {3GPP TR 38.901 version 18.0.0},
  year    = {2024},
  number  = {(Release 18)},
}

@Article{ITU:15,
  author  = {ITU-R},
  title   = {Propagation data and prediction methods for the planning of indoor radiocommunication systems and radio local area networks in the frequency range 300 {MHz} to 100 {GHz}},
  journal = {P Series Radiowave propagation, Recommendation ITU-R},
  year    = {2015},
  number  = {1238--8},
}

@InProceedings{Malarski:19,
  author    = {Malarski, Krzysztof Mateusz and Thrane, Jakob and Bech, Markus Greve and Macheta, Kamil and Christiansen, Henrik Lehrmann and Petersen, Martin Nordal and Ruepp, Sarah},
  title     = {Investigation of Deep Indoor NB-IoT Propagation Attenuation},
  booktitle = {2019 IEEE 90th Vehicular Technology Conference (VTC2019-Fall)},
  year      = {2019},
  pages     = {1-5},
  doi       = {10.1109/VTCFall.2019.8891414},
}

@InProceedings{wan:23,
  author       = {Lu Wan and Xiaobing Dai and Torsten Welfonder and Ekaterina Petrova and Pieter Pauwels},
  title        = {Semi-automated thermal envelope model setup for adaptive model predictive control with event-triggered system identification},
  booktitle    = {Building Simulation 2023: 18th Conference of IBPSA},
  year         = {2023},
  volume       = {18},
  series       = {Building Simulation},
  pages        = {3664--3671},
  address      = {Shanghai, China},
  month        = {9},
  publisher    = {IBPSA},
  doi          = {https://doi.org/10.26868/25222708.2023.1611},
  issn         = {2522-2708},
  organisation = {IBPSA},
}

@Article{Tangsunantham:22,
  author         = {Tangsunantham, Natthanan and Pirak, Chaiyod},
  title          = {Experimental Performance Analysis of {Wi-SUN} Channel Modelling Applied to Smart Grid Applications},
  journal        = {Energies},
  year           = {2022},
  volume         = {15},
  number         = {7},
  issn           = {1996-1073},
  article-number = {2417},
  doi            = {10.3390/en15072417},
}

@InProceedings{Hoo:23,
  author    = {Hoo, Filbert and Lim Tan, Forest Su and Ching Bon Chan, Raymond and Waszecki, Peter and Keoh, Sye Loong and Kiat Seow, Chee and Li, Minghui and Cao, Qi and Sum, Chin Sean},
  title     = {5G-Wi-SUN for Building Management System},
  booktitle = {2023 6th International Conference on Applied Computational Intelligence in Information Systems (ACIIS)},
  year      = {2023},
  pages     = {1-6},
  doi       = {10.1109/ACIIS59385.2023.10367377},
}

\end{document}